\documentclass[pra,twocolumn,longbibliography]{revtex4-2}
\usepackage{graphicx}  % needed for figures
\usepackage{dcolumn}   % needed for some tables
\usepackage{bm}        % for math
\usepackage{verbatim}   % for math
\usepackage{graphicx}
\usepackage{epstopdf}
\usepackage{footnote}
\usepackage{epsfig}
\usepackage{subfigure}
\usepackage{amssymb}
\usepackage{amsmath,bm}
\usepackage{xcolor}
\usepackage{float}
\usepackage{subfigure}
\usepackage[most]{tcolorbox}
\usepackage{soul}
\usepackage{color}
\usepackage[colorlinks=true,linkcolor=blue,allcolors=blue]{hyperref}%
\usepackage[framemethod=TikZ]{mdframed}
\usepackage{lipsum}
\mdfdefinestyle{MyFrame}{%
	linecolor=blue,
	outerlinewidth=2pt,
	roundcorner=20pt,
	innertopmargin=\baselineskip,
	innerbottommargin=\baselineskip,
	innerrightmargin=20pt,
	innerleftmargin=20pt,
	backgroundcolor=gray!50!white}
\begin{document}
	
	{PHYSICAL REVIEW APPLIED,~Vol.~,  ~(2026)}
	
	\headsep = 40pt
	\title{Photon-Blockade Analogue Nonreciprocal Absorption in Coherent Spatiotemporal Metasurfaces}
	\author{Sajjad Taravati}
	
	\affiliation{School of Electronics and Computer Science, University of Southampton, Southampton SO17 1BJ, UK}
	\email{e-mail: s.taravati@soton.ac.uk}

\begin{abstract}
Controlling the flow of electromagnetic energy is essential for advancing quantum technologies. We introduce a spatiotemporally modulated superconducting metasurface that exhibits photon-blockade-analogue nonreciprocal absorption. In this system, the frequency of incident radiation is matched to the modulation frequency of the metasurface, enabling one-way directional absorption. Forward-traveling waves undergo resonant coupling to higher-order Floquet harmonics and are absorbed within the slab, while backward-traveling waves transmit freely without interaction. This behavior arises from classical wave interference and harmonic conversion in a space-time periodic medium—a classical analogue of quantum photon blockade. We present a design based on a superconductor-semiconductor metasurface incorporating cascaded Josephson field-effect transistors (JoFETs) for millikelvin-temperature operation. 
Starting from the microscopic Hamiltonian of a single gate-tunable JoFET cell, we derive the system's classical circuit relations, effective space-time-periodic permeability, Floquet band structure, and isofrequency diagrams from first principles, and validate the resulting nonreciprocal absorption with full-wave simulations. These findings establish a pathway toward compact, nonreciprocal superconducting devices for quantum information processing and microwave photonics.
\end{abstract}
	
\maketitle

\section{Introduction}
Photon blockade is a quantum optical phenomenon where the absorption or transmission of a single photon inhibits the absorption or transmission of subsequent photons. This effect, typically observed in systems such as atoms, quantum dots, or superconducting circuits coupled to resonant cavities, involves the system entering an excited state after the first photon interaction. This state shift prevents further photon interactions until the system returns to its ground state, which is crucial for quantum information processing and communication. Historically, photon blockade has been achieved in systems with strong coupling between quantum emitters and optical cavities, where the nonlinearity creates an energy gap sufficient to block additional photons~\cite{birnbaum2005photon, chakram2022multimode}. Recent advancements have expanded this concept to nonreciprocal photon blockade in rotating nonlinear devices, where device geometry or external fields control the directionality of photon interactions~\cite{huang2018nonreciprocal}. Additionally, loss-induced nonreciprocal photon blockade has been demonstrated, where engineered losses enable nonreciprocal photon dynamics, enhancing control over photon flow in quantum circuits \cite{vrajitoarea2020quantum,li2024loss}.

Quantum nonreciprocity can be achieved using chirality~\cite{zhang2025chirality}, nonlinearity~\cite{Rosario}, giant magnon molecule~\cite{wang2026quantum}, superconducting nonreciprocity in a solid-state platform~\cite{ando2020observation}, and synthetic gauge fields~\cite{barzanjeh2025nonreciprocity}. Nonreciprocal elements like isolators and circulators are critical for protecting qubits from back-reflected signals and routing quantum information. Recent studies have explored various mechanisms for achieving superconducting quantum nonreciprocity, including nonreciprocal quantum entanglement between frequency distinct qubits~\cite{taravati2025_entangle}, dynamic parametric interactions~\cite{wang2023quantum,wanjura2023quadrature}, and frequency-multiplexed millimeter-wave qubits via nonreciprocal control bus~\cite{taravati2025frequency}. Observations of superconducting nonreciprocity through diamagnetic effects and engineered circuit asymmetries offer robust platforms for nonreciprocal quantum devices \cite{ando2020observation,zhang2020nonreciprocal,sundaresh2023diamagnetic,xiong2024electrical}. Traditional directional absorbers and electronic components, such as varactors, transistors, and diodes, face limitations and introduce noise at millikelvin temperatures typical of superconducting quantum technologies. To overcome these challenges, we propose a spatiotemporally modulated superconducting metasurface that exhibits photon-blockade-analogue nonreciprocal absorption. By leveraging the unique properties of superconducting materials and dynamically modulating the metasurface structure, we achieve directional absorption where incident waves are selectively absorbed or transmitted based on their direction of incidence. This classical analogue of photon blockade, where forward-traveling waves undergo resonant harmonic conversion and absorption while backward-traveling waves transmit freely, offers new capabilities for controlling electromagnetic wave propagation in quantum-compatible platforms. The approach presents significant opportunities for advancing nonreciprocal devices and photon management in millikelvin-temperature quantum systems.

Our approach involves nonlinear space-time-modulated metasurfaces incorporating gate-controlled Josephson field-effect transistors (JoFETs)~\cite{mayer2020gate,o2021epitaxial,phan2023gate}. These metasurfaces deliver highly efficient directional absorption while maintaining superior performance in superconducting quantum applications~\cite{devoret2013superconducting}. By enabling the miniaturization and integration of directional absorption components, our method preserves quantum state integrity and represents a significant leap forward in superconducting quantum technologies. The metasurface, featuring cascaded space-time-varying JoFETs, exploits the quantum mechanical phenomenon of supercurrent flow without resistance~\cite{makhlin2001quantum,kleiner2021space}. This design harnesses the nonlinearity of JoFETs and their spatial and temporal modulation capabilities to advance directional absorption and wave reflection. Our approach achieves efficient directional absorption with simultaneous amplification, a rare combination in conventional devices. It addresses inefficiencies and spurious signal generation, paving the way for high-performance signal processing applications.

Furthermore, our metasurface demonstrates nonreciprocal behavior, with different absorption responses for incident waves from opposing directions, emphasizing the unidirectional properties imparted by the spatiotemporal modulation \cite{Taravati_PRB_SB_2017,Bahl_2018time,Taravati_Kishk_MicMag_2019,Taravati_NC_2021,taravati2025designing} and time-periodic devices\cite{zhang2018space,Taravati_Kishk_PRB_2018,Grbic2019serrodyne,taravati2020full,cardin2020surface,Taravati_AMA_PRApp_2020,Taravati_ACSP_2022,taravati2025light,sisler2024electrically,taravati2026temporal,taravati2026freqBragg}. Utilizing analytical techniques such as Bloch-Floquet solutions, we provide a comprehensive understanding of the wave scattering phenomena from our metasurface, validating its design and offering insights for further optimization. Unlike linear systems constrained by phase-matching and dispersion, our nonlinear superconducting spatiotemporal media achieve superior performance in a compact form, making them ideal for millikelvin-temperature superconducting quantum technologies.

%%%%%%%%%%%%%%%%%%%%%%%%%%%%%%%%%%%%%%%%%%%%%%%%%%%%%%%%%%%%%%%%%%%%%%%%%%%%%

\section{Photon-Blockade Analogue Mechanism for Nonreciprocal Absorption}\label{sec:Theory}

\begin{figure}
	\begin{center}
		\subfigure[]{\label{Fig:energy_a}
			\includegraphics[width=1\linewidth]{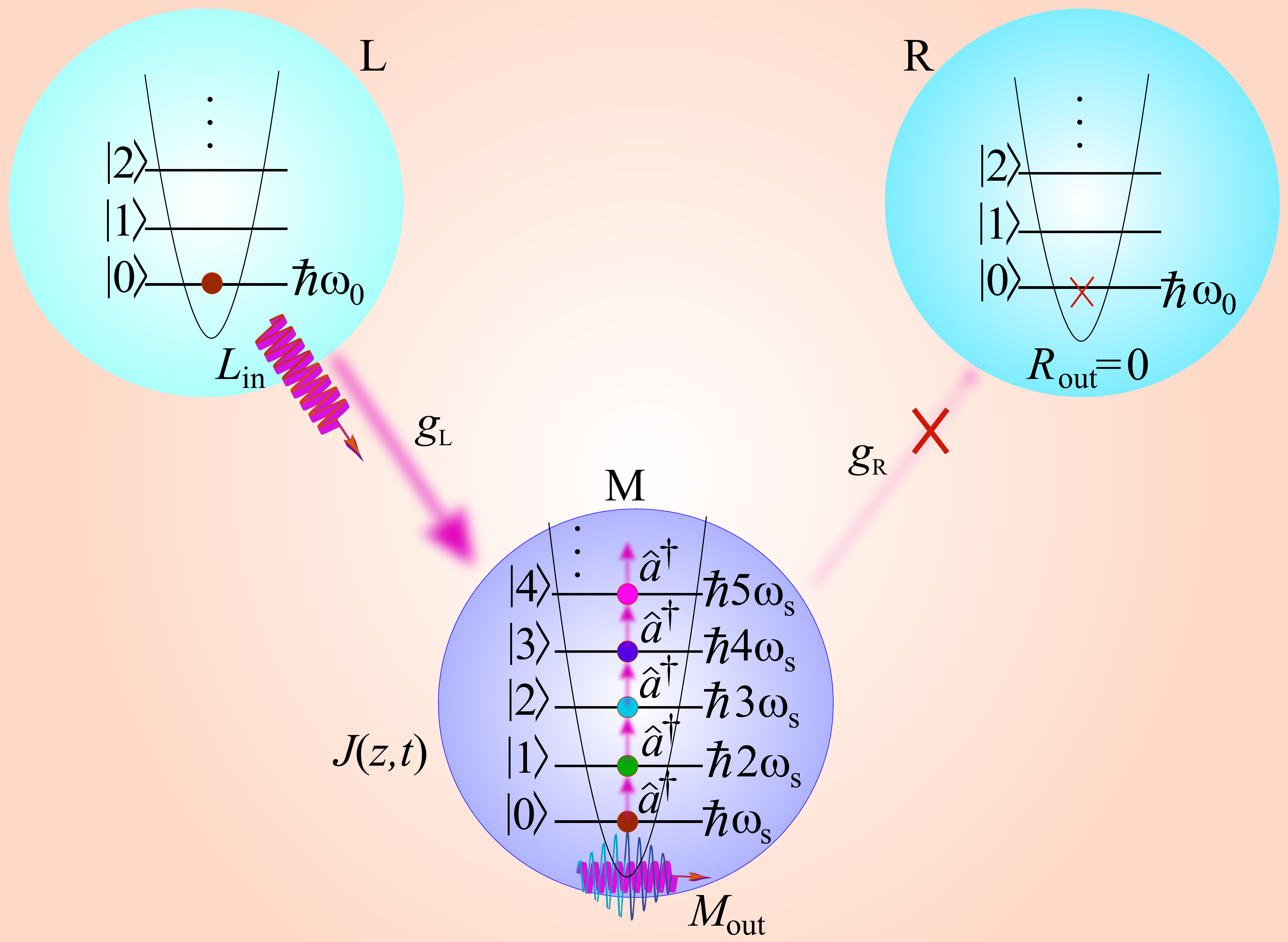}  }
		\subfigure[]{\label{Fig:energy_b}
			\includegraphics[width=1\linewidth]{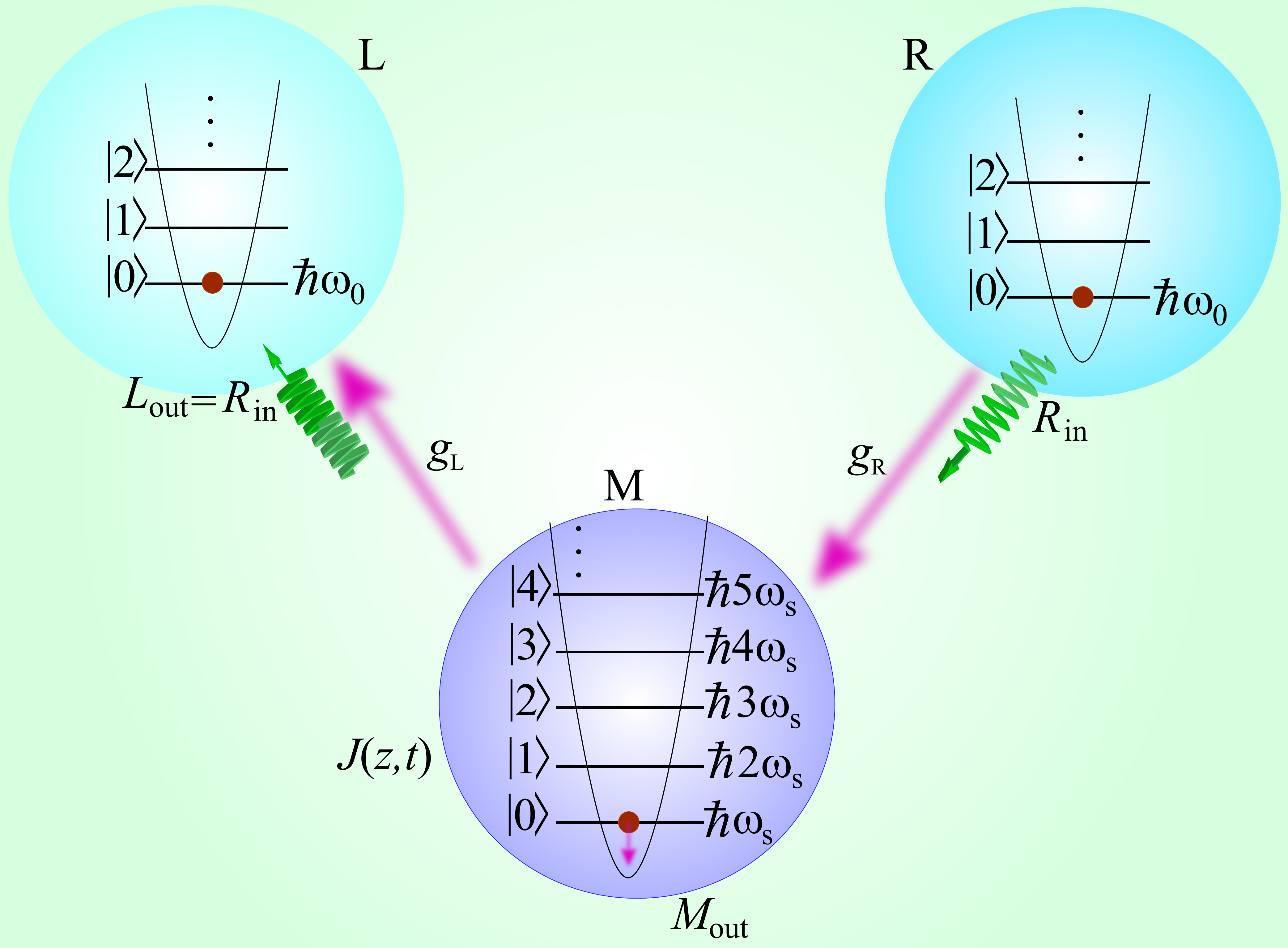}  }
		\caption{Nonreciprocal absorption in superconducting spatiotemporal metasurfaces, where the space-time modulation and incident wave share the same frequency ($\omega_\text{s}=\omega_0$). The system comprises space-time-modulated nonlinear unit cells coupled together. a)~Incidence from the left port ($L_\text{in}$): Energy transition to higher states occur due to the strong interaction and one-way coherency with the nonlinear left-to-right traveling space-time modulation, resulting in zero transmission to the right port ($R_\text{out}=0$). b)  Illustration of the physical mechanism for incidence from the right port ($R_\text{in}$): Free passage of wave to the left port ($L_\text{out}=R_\text{in}$) occurs due to the lack of transition to higher energy states and weak interaction with the opposite-direction traveling nonlinear space-time modulation, resulting in full transmission of waves.}
		\label{Fig:1}
	\end{center}
\end{figure}

Consider waves with frequency $\omega_0$ incident on a superconducting slab with a space-time periodic current density $J(z,t)$, modulated at a temporal frequency $\omega_\text{s}$, where $\omega_\text{s} = \omega_0$. We demonstrate that the temporal coherence between the incident wave and the periodic modulation of the superconducting slab induces a nonreciprocal transmission. This phenomenon arises because the energy associated with the space-time modulation effectively inhibits the transmission of subsequent photons. Instead, the incident photons transition to higher energy states within the slab and are absorbed, preventing their exit. This results in a one-way quantum absorption process, facilitated by the coherent spatiotemporal modulation of the superconducting slab.

Figure~\ref{Fig:1} illustrates the principle of spatiotemporal nonreciprocal absorption by leveraging the interplay between the incident photons and the one-way coherent dynamic modulation. Figure~\ref{Fig:energy_a} demonstrates the photon blockade effect when photons are incident from the left port ($L_\text{in}$). Here, the photons interact strongly with the space-time modulation traveling from left to right. This interaction causes the system to enter an excited state, leading to energy transitions and blocking the transmission of photons to the right port ($R_\text{out}=0$). In contrast, Fig.~\ref{Fig:energy_b} illustrates the scenario for wave incident from the right port ($R_\text{in}$). In this case, the interaction with the space-time modulation traveling in the opposite direction is weak. Consequently, there is minimal energy transition and the wave is transmitted freely to the left port ($L_\text{out}=R_\text{in}$). This nonreciprocal behavior, where the wave transmission and absorption are direction-dependent, underscores the potential of spatiotemporal metasurfaces for applications in quantum communication and sensing.

\begin{figure}
	\begin{center}
		\includegraphics[width=1\linewidth]{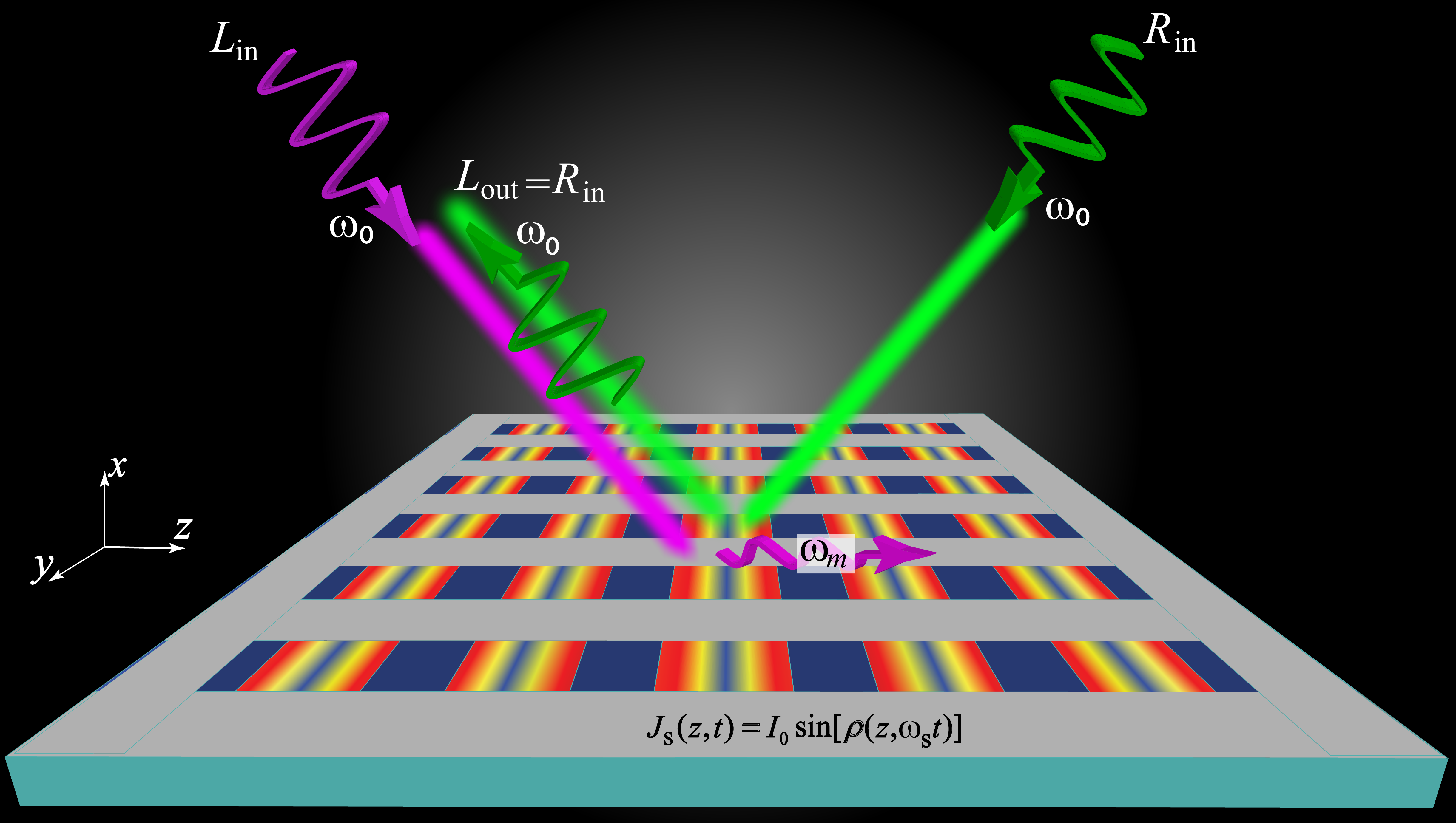} 
		\caption{One-way absorption in superconductor-semiconductor quantum spatiotemporal metasurface.}
		\label{Fig:sch}
	\end{center}
\end{figure}

\begin{figure}
	\begin{center}
			\includegraphics[width=1\linewidth]{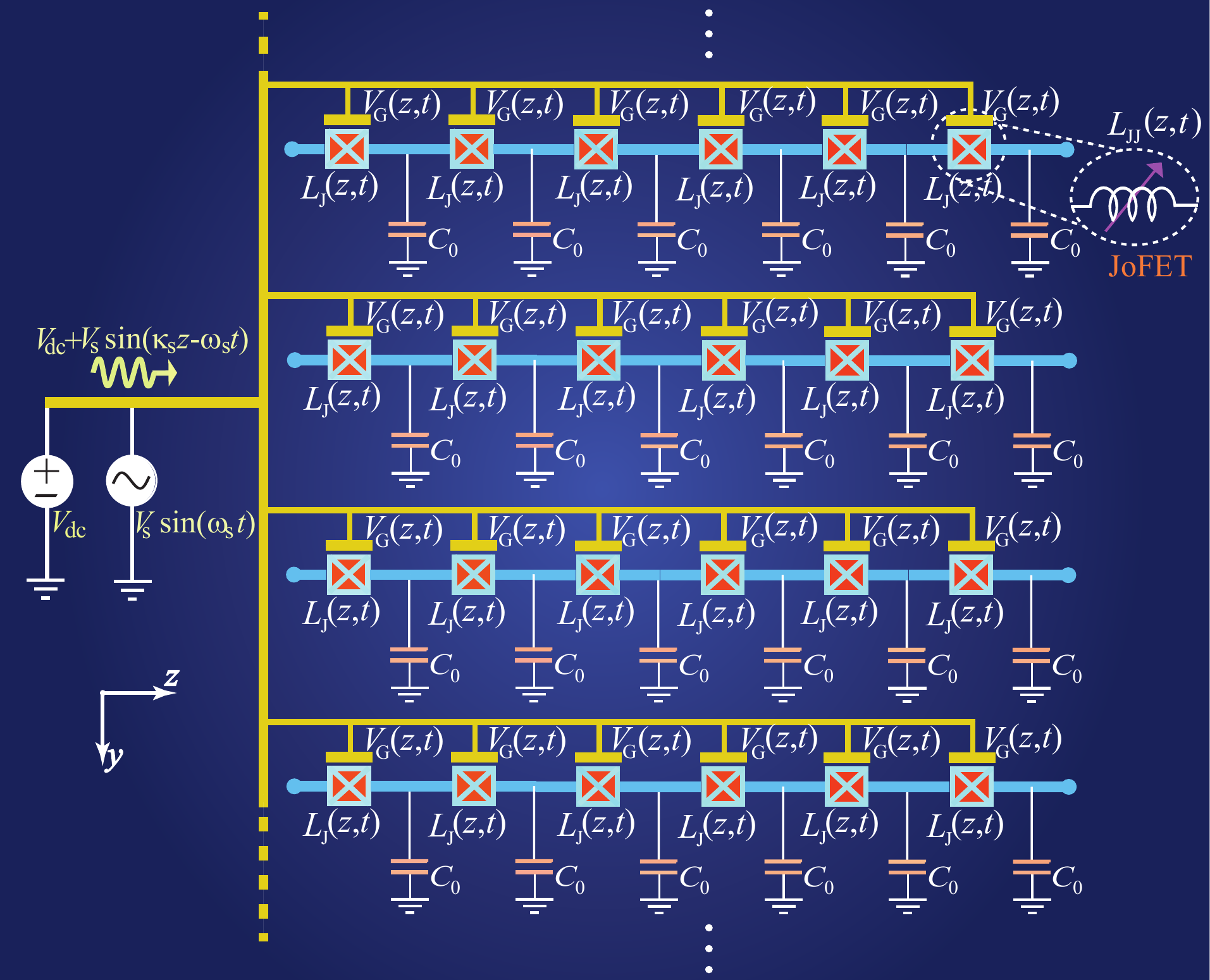} 
		\caption{Circuit model of the metasurface, consisting of an array of spatiotemporally-modulated gate-controlled Josephson field-effect transistors (JoFETs).}
		\label{Fig:equiv}
	\end{center}
\end{figure}

Figure~\ref{Fig:sch} illustrates the proposed one-way absorption mechanism facilitated by the quantum spatiotemporal metasurface and its functionality. The metasurface operates as a nonreciprocal element, permitting transmission in one direction while blocking it in the opposite direction. Figure~\ref{Fig:equiv} presents an equivalent circuit model of the metasurface, which comprises an array of space-time-periodic gate-controlled Josephson
field-effect transistors (JoFETs). This circuit model is essential for
understanding the underlying physical processes governing the
nonreciprocal behavior. We derive its governing equations, the
spatiotemporally modulated current, and the resulting effective
permeability of the array from first principles in Sec.~III, starting
from the microscopic Hamiltonian of a single JoFET cell.

\section{Theoretical Analysis}
\subsection{Hamiltonian derivation}
We build the theoretical model from first principles, starting from the
microscopic quantum circuit Hamiltonian of a single JoFET cell and deriving
the classical circuit relations, the effective-medium description, and the
Floquet space-time analysis as direct consequences. For the $j$-th JoFET cell, the Hamiltonian comprises the charging energy of
the junction capacitance $C_j$ and the gate-tunable Josephson coupling
energy,
\begin{equation}
	H_j = 4E_C n_j^2 - E_J(V_g,t)\cos\phi_j,
	\label{eq:Hj}
\end{equation}
where $n_j$ is the number of Cooper pairs on the junction ($Q_j=2en_j$),
$\phi_j$ is the superconducting phase difference across the junction,
$E_C=e^2/(2C_j)$ is the single-electron charging energy, and
$E_J(V_g,t)=\Phi_0 I_c(V_g,t)/(2\pi)$ is the gate-voltage-dependent
Josephson coupling energy, with $\Phi_0=h/2e$ the magnetic flux quantum.
The phase and Cooper-pair number are canonically conjugate,
$[\hat\phi_j,\hat n_j]=i$.

\textbf{Josephson relations from the Heisenberg equations of motion.}
Rather than postulating the current-phase and voltage-phase relations, we
derive them directly from $H_j$. For the phase,
\begin{equation}
	\dot{\hat\phi}_j = \frac{i}{\hbar}[H_j,\hat\phi_j]
	= \frac{8E_C}{\hbar}\hat n_j = \frac{2\pi}{\Phi_0}\hat V_j,
	\label{eq:heisenberg_V}
\end{equation}
where $V_j=Q_j/C_j$, so that $V_j=(\Phi_0/2\pi)\,d\phi_j/dt$. For the
Cooper-pair number,
\begin{equation}
	\dot{\hat n}_j = \frac{i}{\hbar}[H_j,\hat n_j]
	= -\frac{E_J}{\hbar}\sin\hat\phi_j,
	\label{eq:heisenberg_I}
\end{equation}
so that the junction current $I_j=2e\,\dot n_j$ becomes
$I_j = (2\pi E_J/\Phi_0)\sin\phi_j = I_c(V_g)\sin\phi_j$. Equations
\eqref{eq:heisenberg_V}--\eqref{eq:heisenberg_I} are the familiar Josephson
relations, now obtained as a direct consequence of the microscopic
Hamiltonian~\eqref{eq:Hj}, rather than as an independent starting postulate.

\textbf{Spatiotemporal gate drive.} In the proposed metasurface, the gate
voltage is not applied uniformly: it is injected as a signal that
propagates along the array (Fig.~3) and reaches the $j$-th cell, located
at position $z$, with a propagation-induced delay,
\begin{equation}
	V_g(z,t) = V_\text{dc} + V_\text{rf}\sin(\kappa_s z-\omega_s t+\phi).
	\label{eq:Vg}
\end{equation}
Experimental studies of Al/InAs JoFETs of the type used here~\cite{mayer2020gate}
show that gate voltage primarily induces an anomalous phase shift in the
current-phase relation, with comparatively weak modulation of the
critical-current amplitude. We therefore treat $I_0$ as constant and let
the gate drive act on the Josephson phase response through the relation
just derived in Eq.~\eqref{eq:heisenberg_V}:
\begin{equation}
	\rho_g(z,t) \equiv \frac{2\pi}{\Phi_0}\int_{-\infty}^t V_g(z,t')\,dt'
	= \widetilde\Phi_\text{dc} + \widetilde\Phi_\text{rf}\sin[\kappa_s z-\omega_s t+\phi],
	\label{eq:rho_g}
\end{equation}
where $\widetilde\Phi_\text{dc}=2\pi\Phi_\text{dc}/\Phi_0$ and
$\widetilde\Phi_\text{rf}=2\pi\Phi_\text{rf}/\Phi_0$ are the normalized
DC and RF fluxes, with $\Phi_\text{rf}=V_\text{rf}/\omega_s$. The
gate-modulated Josephson coupling energy is thus obtained directly from
Eq.~\eqref{eq:Hj}--\eqref{eq:rho_g} as
\begin{equation}
\begin{split}
	E_J(z,t) &= \frac{\Phi_0 I_0}{2\pi}\cos\!\big[\rho_g(z,t)\big]\\&
	= \frac{\Phi_0 I_0}{2\pi}\cos\!\left(\widetilde\Phi_\text{dc}+\widetilde\Phi_\text{rf}\sin[\kappa_s z-\omega_s t+\phi]\right).	
\end{split}
	\label{eq:EJ_zt}
\end{equation}

\textbf{Discrete and continuum Hamiltonian.} Using
$Q_j=C_j(\Phi_0/2\pi)\dot\phi_j$ [Eq.~\eqref{eq:heisenberg_V}], the discrete
Hamiltonian of the $j$-th cell reads
\begin{equation}
	H_j = \frac{1}{2}C_j\left(\frac{\Phi_0}{2\pi}\right)^2\left(\frac{\partial\phi_j}{\partial t}\right)^2 - E_{J,j}\cos\phi_j.
	\label{eq:Hj_discrete}
\end{equation}
The junctions are coupled via the shared inductance $L_s(z,t)$ between
them, contributing an inductive energy
\begin{equation}
	H_{\text{ind},j} = \frac{1}{2L_s(z,t)}\left(\frac{\Phi_0}{2\pi}\right)^2(\phi_{j+1}-\phi_j)^2.
	\label{eq:Hind_j}
\end{equation}
Taking the continuum limit ($\Delta z\to0$, $\phi_{j+1}-\phi_j\approx\Delta z\,\partial\phi/\partial z$, $\sum_j\Delta z\to\int dz$), the total Hamiltonian becomes
\begin{equation}
\begin{split}	
	H &= \int dz\Big[\frac{1}{2}C(z)\left(\frac{\Phi_0}{2\pi}\right)^2\left(\frac{\partial\phi}{\partial t}\right)^2\\&
	+ \frac{1}{2L_s(z,t)}\left(\frac{\Phi_0}{2\pi}\right)^2\left(\frac{\partial\phi}{\partial z}\right)^2
	- E_J(z,t)\cos\phi\Big],
\end{split}
	\label{eq:H_continuum}
\end{equation}
with Hamiltonian density
\begin{equation}
\begin{split}	
	&\mathcal H(z,t) = \frac{1}{2}C(z)\left(\frac{\Phi_0}{2\pi}\right)^2\left(\frac{\partial\phi}{\partial t}\right)^2
	\\&\quad+ \frac{1}{2L_s(z,t)}\left(\frac{\Phi_0}{2\pi}\right)^2\left(\frac{\partial\phi}{\partial z}\right)^2
	- E_J(z,t)\cos\phi.
\end{split}
	\label{eq:H_density}
\end{equation}

\textbf{Quantization.} Elevating $\phi$ to an operator-valued field and
expanding in plane-wave modes,
\begin{equation}
	\hat\phi(z,t) = \sum_k\sqrt{\frac{\hbar}{2\omega_k C_k}}\left(\hat a_k e^{i(kz-\omega_k t)}+\hat a_k^\dagger e^{-i(kz-\omega_k t)}\right),
	\label{eq:phi_quantization}
\end{equation}
the quadratic (free-field) part of the Hamiltonian becomes
$H_0=\sum_k\hbar\omega_k \hat a_k^\dagger\hat a_k$, with dispersion
$\omega_k=|k|/\sqrt{L_sC}$. For small phase fluctuations, expanding
$\cos\hat\phi=1-\hat\phi^2/2+\hat\phi^4/24+\mathcal O(\hat\phi^6)$, the
quartic term generates the nonlinear interaction responsible for harmonic
generation:
\begin{equation}
	H_\text{int} = \frac{1}{24}\int dz\,E_J(z,t)\,\hat\phi^4(z,t).
	\label{eq:Hint_quartic}
\end{equation}

\textbf{Fourier expansion of $E_J(z,t)$.} Since $E_J(z,t)$
[Eq.~\eqref{eq:EJ_zt}] is periodic in $\psi\equiv\kappa_s z-\omega_s t+\phi$,
expanding $\cos(\widetilde\Phi_\text{rf}\sin\psi)$ and
$\sin(\widetilde\Phi_\text{rf}\sin\psi)$ in Taylor series (as in the
Appendix) yields the Fourier series
\begin{equation}
	E_J(z,t) = \sum_{m=-\infty}^{\infty} E_{J,m}\,e^{-im(\kappa_s z-\omega_s t+\phi)},
	\label{eq:EJ_fourier}
\end{equation}
with coefficients $E_{J,m}$ given explicitly in terms of
$\widetilde\Phi_\text{dc}$, $\widetilde\Phi_\text{rf}$ by the same
expressions as Eq.~(2d) of the original manuscript, with $I_\mu$ there
replaced by $\Phi_0 I_0/2\pi$.

Substituting Eqs.~\eqref{eq:phi_quantization} and \eqref{eq:EJ_fourier}
into Eq.~\eqref{eq:Hint_quartic}, and retaining, in the rotating-wave
approximation, only the terms coupling the fundamental mode to the
modulation harmonics, the interaction Hamiltonian becomes
\begin{equation}
	H_\text{int} = \sum_{m,n} g_{m,n}\left(\hat a_n^\dagger \hat a_{n+m}\hat b_m + \text{h.c.}\right),
	\label{eq:Hint_final}
\end{equation}
where $g_{m,n}\propto E_{J,m}$ and $\hat b_m$ is the modulation-harmonic
operator. Combining $H_0$, $H_\text{int}$, and the energy of the
modulation harmonics themselves,
\begin{equation}
	\begin{split}
		\mathcal{H} = &\sum_{n,k} \hbar \omega_k \hat{a}_{n,k}^\dagger \hat{a}_{n,k} + \sum_{m} \hbar \omega_m \hat{b}_m^\dagger \hat{b}_m \\&+ \sum_{m,n} g_{m,n} \left(\hat{a}_{n}^\dagger \hat{a}_{n+m} \hat{b}_m + \text{h.c.}\right),
		\label{eq:H_total_modes}
	\end{split}
\end{equation}
In the strong-drive regime, treating $\hat b_m\to\beta_m e^{-i\omega_s t}$
as a classical coherent state and retaining only coupling to the first
harmonic, the effective Hamiltonian simplifies to
\begin{equation}
	\begin{split}
		&\mathcal{H}_{\text{eff}} = \hbar\omega \hat{a}^\dagger \hat{a} + \sum_m \hbar\omega_m \hat{a}_m^\dagger \hat{a}_m \\&\quad+ g\left(\hat{a}^\dagger \hat{a}_m e^{-i(\kappa_s z - \omega_\text{s} t + \phi)} + \hat{a}_m^\dagger \hat{a} e^{i(\kappa_s z - \omega_\text{s} t + \phi)}\right). 
	\end{split}
\end{equation}
The Heisenberg equation of motion for $\hat a_n$ then gives
\begin{equation}
	i\hbar\frac{d\hat a_n}{dt} = \hbar\omega_n\hat a_n + \sum_m g_{m,n}\hat a_{n+m}e^{-i(\kappa_s z-\omega_s t+\phi)},
	\label{eq:heisenberg_an}
\end{equation}
connecting the quantized description to the classical wave equation used
in Sec.~III.C.

\subsection{Space-time-periodic effective permeability}
Having established the microscopic origin of the gate-modulated Josephson
coupling $E_J(z,t)$ [Eq.~\eqref{eq:EJ_zt}], we now derive the classical
effective-medium description of the array used in the dispersion analysis
of Sec.~III.C.

Linearizing the current-phase relation of Eq.~\eqref{eq:heisenberg_I} at
the gate-driven operating point $\rho_g(z,t)$ gives
$dJ/dt=(2\pi I_0/\Phi_0)\cos[\rho_g(z,t)]\,V(z,t)$, i.e.,
$V(z,t)=L_s(z,t)\,dJ/dt$ with the shared inductance already introduced in
Eq.~\eqref{eq:Hind_j} given, via Eq.~\eqref{eq:EJ_zt}, by
\begin{equation}
	L_s(z,t) = \left(\frac{\Phi_0}{2\pi}\right)^2\frac{1}{E_J(z,t)}
	= \frac{\Phi_0}{2\pi I_0\cos[\rho_g(z,t)]}.
	\label{eq:Ls_relation}
\end{equation}
The array's effective magnetic permeability follows directly as
\begin{equation}
	\begin{split}
&	\mu_s(z,t) = \frac{l\,L_s(z,t)}{\mu_0 A}\\&\quad
	= \frac{l}{\mu_0 A}\frac{\Phi_0}{2\pi I_0}\sec\!\left(\widetilde\Phi_\text{dc}+\widetilde\Phi_\text{rf}\sin[\kappa_s z-\omega_s t+\phi]\right),
	\end{split}
	\label{eq:mu_s}
\end{equation}
and the corresponding Josephson current density of the array reads
\begin{equation}
	J(z,t) = I_0\sin\!\big[\rho_g(z,t)\big].
	\label{eq:J_current}
\end{equation}
As $\mu_s(z,t)$ is periodic in both space and time, its inverse
$a(z,t)=1/\mu_s(z,t)$ admits the Fourier expansion
\begin{equation}
	a(z,t) = \sum_{m=-\infty}^{\infty} a_m\,e^{-jm(\kappa_s z-\omega_s t+\phi)}.
	\label{eq:a_zt}
\end{equation}
Since $a(z,t)=(2\pi\mu_0 I_0 A/\Phi_0 l)\cos[\rho_g(z,t)]$ is directly
proportional to $E_J(z,t)$ already expanded in Eq.~\eqref{eq:EJ_fourier},
the Fourier coefficients are related by a single constant,
\begin{equation}
	a_m = \frac{4\pi^2\mu_0 A}{\Phi_0^2 l}\,E_{J,m},
	\label{eq:am_relation}
\end{equation}
so the coefficients $a_m$ used in the dispersion analysis of Sec.~III.C
are simply those already derived for $E_{J,m}$ in Eq.~\eqref{eq:EJ_fourier},
rescaled by this constant factor.

As the nonlinear permeability of JoFETs exhibits periodicity in both space and time, its inverse denoted as $a(z,t)=  1/\mu_\text{s}(z,t)$ can be represented by a Fourier series expansion, i.e.,
\begin{subequations}
\begin{equation}\label{eqa:perm0}
		a(z,t)=  \sum_{m=-\infty}^{+\infty} a_m e^{-jm(\kappa_\text{s} z-\omega_\text{s} t+\phi)},
\end{equation}
where $a_m$ are the Fourier coefficients of the expansion. We define $I_{\mu}=2\pi \mu_0 I_0 A/(\Phi_0 l)$ and $\psi=\kappa_\text{s} z-\omega_\text{s} t+\phi$, and expand the cosine term using the Taylor's series expansion, as
\begin{equation}
		\begin{split}
			\cos&\left( \widetilde{\Phi}_\text{rf} \sin \psi\right)=
			\left(1-\dfrac{{\widetilde{\Phi}_\text{rf}^2}}{4}+\dfrac{{\widetilde{\Phi}_\text{rf}^4}}{64}-\dfrac{{\widetilde{\Phi}_\text{rf}^6}}{2304} +\dfrac{{\widetilde{\Phi}_\text{rf}^8}}{147456}\right)\\&+\cos(2\psi) \left( \dfrac{{\widetilde{\Phi}_\text{rf}^2}}{4}- \dfrac{{\widetilde{\Phi}_\text{rf}^4}}{48}  +\dfrac{{\widetilde{\Phi}_\text{rf}^6}}{1536} - \dfrac{{\widetilde{\Phi}_\text{rf}^8}}{92160}\right) \\& \qquad+\cos(4\psi)
			\left(  \dfrac{{\widetilde{\Phi}_\text{rf}^4}}{192}  - \dfrac{{\widetilde{\Phi}_\text{rf}^6}}{3840} + \dfrac{{\widetilde{\Phi}_\text{rf}^8}}{184320}\right) \\&
			+\cos(6\psi)
			\left(  \dfrac{{\widetilde{\Phi}_\text{rf}^6}}{23040} - \dfrac{{\widetilde{\Phi}_\text{rf}^8}}{645120}\right)
			+\cos(8\psi)
			\dfrac{{\widetilde{\Phi}_\text{rf}^8}}{5160960},
		\end{split}
	\end{equation}
	\begin{equation}
		\begin{split}
			\sin&\left( \widetilde{\Phi}_\text{rf} \sin \psi\right)=\sin\psi \left(\widetilde{\Phi}_\text{rf}-\dfrac{{\widetilde{\Phi}_\text{rf}^3}}{8}+\dfrac{{\widetilde{\Phi}_\text{rf}^5}}{192}-\dfrac{{\widetilde{\Phi}_\text{rf}^7}}{9216}  \right)   \\&
			+\sin(3\psi) \left( \dfrac{{\widetilde{\Phi}_\text{rf}^3}}{24}-\dfrac{{\widetilde{\Phi}_\text{rf}^5}}{384}+\dfrac{{\widetilde{\Phi}_\text{rf}^7}}{15360}  \right) \\&
			+\sin(5\psi) \left( \dfrac{{\widetilde{\Phi}_\text{rf}^5}}{1920}-\dfrac{{\widetilde{\Phi}_\text{rf}^7}}{46080}  \right)
			-\sin(7\psi)  \dfrac{{\widetilde{\Phi}_\text{rf}^7}}{322560},
		\end{split}
	\end{equation}
	where 
	\begin{equation}\label{eqa:a_n}
		\begin{split}
			&	a_0=I_{\mu} \cos\left( \widetilde{\Phi}_\text{dc} \right) 	\left(1-\dfrac{{\widetilde{\Phi}_\text{rf}^2}}{4}+\dfrac{{\widetilde{\Phi}_\text{rf}^4}}{64}-\dfrac{{\widetilde{\Phi}_\text{rf}^6}}{2304} +\dfrac{{\widetilde{\Phi}_\text{rf}^8}}{147456}\right); \quad
		\\&	a_1=-a_{-1}=\dfrac{I_{\mu} }{2j }\sin\left( \widetilde{\Phi}_\text{dc} \right) \left(\widetilde{\Phi}_\text{rf}-\dfrac{{\widetilde{\Phi}_\text{rf}^3}}{8}+\dfrac{{\widetilde{\Phi}_\text{rf}^5}}{192}-\dfrac{{\widetilde{\Phi}_\text{rf}^7}}{9216}  \right)  \\&
			a_2=a_{-2}= \dfrac{I_{\mu}}{8}  \cos\left( \widetilde{\Phi}_\text{dc} \right) \left( {\widetilde{\Phi}_\text{rf}^2}- \dfrac{{\widetilde{\Phi}_\text{rf}^4}}{12}  +\dfrac{{\widetilde{\Phi}_\text{rf}^6}}{384} - \dfrac{{\widetilde{\Phi}_\text{rf}^8}}{23040}\right	); \quad
		\\&	a_3=-a_{-3}=\dfrac{I_{\mu}}{j48 }\sin\left( \widetilde{\Phi}_\text{dc} \right) 
			\left( {\widetilde{\Phi}_\text{rf}^3}-\dfrac{{\widetilde{\Phi}_\text{rf}^5}}{16}+\dfrac{{\widetilde{\Phi}_\text{rf}^7}}{640}  \right); \\&
			a_4=a_{-4}= \dfrac{I_{\mu}}{384 }  \cos\left( \widetilde{\Phi}_\text{dc} \right)
			\left(  {\widetilde{\Phi}_\text{rf}^4}  - \dfrac{{\widetilde{\Phi}_\text{rf}^6}}{20} + \dfrac{{\widetilde{\Phi}_\text{rf}^8}}{960}\right) ; \quad
		\\&	a_5=-a_{-5}=\dfrac{I_{\mu}}{j3840 }\sin\left( \widetilde{\Phi}_\text{dc} \right) \left( {\widetilde{\Phi}_\text{rf}^5}-\dfrac{{\widetilde{\Phi}_\text{rf}^7}}{24}  \right);\\&
			a_6=a_{-6}= \dfrac{I_{\mu}}{46080 }  \cos\left( \widetilde{\Phi}_\text{dc} \right)
			\left(  {\widetilde{\Phi}_\text{rf}^6} - \dfrac{{\widetilde{\Phi}_\text{rf}^8}}{28}\right) ;
		\\&	a_7=-a_{-7}=\dfrac{I_{\mu}}{j645120 }\sin\left( \widetilde{\Phi}_\text{dc} \right) {\widetilde{\Phi}_\text{rf}^7}; 
		\\&	a_8=a_{-8}= \dfrac{I_{\mu}}{10321920 }  \cos\left( \widetilde{\Phi}_\text{dc} \right)
			{\widetilde{\Phi}_\text{rf}^8}.	
		\end{split}
	\end{equation}
\end{subequations}

\subsection{Floquet space-time harmonics }
Given the spatiotemporal periodicity of the metasurface, the electric and magnetic fields within the slab can be expanded into Floquet space-time harmonics. The magnetic field can be expressed as
\begin{subequations}
	\begin{equation}
		\begin{split}
			\mathbf{H}_\text{s}(x,z,t)&=\mathbf{\hat{y}}\sum_{n }   \varPsi_{n}  e^{ - j \left[ \kappa_{x,n} x+ \kappa_{z,n} z -\omega_n t\right]} ,
		\end{split}
		\label{eqa:H_mod}
	\end{equation}
	and the corresponding electric field is given by
	\begin{equation}
		\begin{split}
			&\mathbf{E}_\text{s}(x,z,t)=
			-\eta_2 \left[\mathbf{\hat{k}}_{\text{s},n}   \times \mathbf{H}_\text{s} (x,z,t)\right]=\\&\eta_2 \sum_{n }   \left(\mathbf{\hat{x}} \frac{\kappa_{z,n} }{ k_n } - \mathbf{\hat{z}}  \frac{\kappa_{x,n} }{ k_n } \right)   \varPsi_{n}  e^{ - j \big[\kappa_{x,n} x+ \kappa_{z,n}z -\omega_n t\big]},
		\end{split}
		\label{eqa:E_mod}
	\end{equation}
with the unit wavevector $\mathbf{\hat{k}}_{\text{s},n} = \mathbf{\hat{x}} \kappa_{x,n}/ k_n + \mathbf{\hat{z}} \kappa_{z,n} / k_n$, where
\begin{equation}
	k_{n} = \sqrt{\kappa_{x,n}^2+\kappa_{z,n}^2 } ,
\end{equation}
is the magnitude of the wavevector of the $  n  $-th harmonic, and 
\begin{equation}
	\kappa_{z,n} = k_0 \sin(\theta_\text{i}) + n \kappa_\text{s},
\end{equation}
\end{subequations}
is the $z$-component of the wavenumber for the $n$th space-time harmonic achieved by phase matching at the interface etween the metasurface and free-space. Additionally, $\kappa_{x,n} = k_n \cos(\theta_n)$ is the $x$-component of the wavenumber (which will be determined through satisfying the dispersion relation). The source-free wave equation for the system is
\begin{subequations}
\begin{equation}
	\nabla^2 \mathbf{H}_\text{s}(x,z,t) - \frac{1}{{{c^2}}}\frac{{{\partial ^2} \left[\mu_\text{s} (z,t)\mathbf{H}_\text{s}(x,z,t) \right]}}{{\partial {t^2}}}=0.
	\label{eqa:wave_eq}
\end{equation}
\indent  Inserting the magnetic field expression from Eq.~\eqref{eqa:H_mod} into the wave equation~\eqref{eqa:wave_eq} yields
\begin{equation}
	\begin{split}
		&\sum_{n}(\kappa_{x,n}^2+ \kappa_{z,n}^2)	   \varPsi_{n}  e^{ - j \left[ \kappa_{x,n} x+ \kappa_{n} z -\omega_n t\right]}\\& -\dfrac{1}{c^2} \dfrac{\partial ^2}{\partial {t^2}} \sum_{m,n}  \widetilde{\mu}_m \varPsi_{m+n}  e^{ - j \left[ \kappa_{x,n} x+ \kappa_{n} z -\omega_n t+\phi\right]} =0,
	\end{split}
	\label{eqa:mmmm}
\end{equation}
which may be cast as
\begin{equation}
	\varPsi_{n} e^{j\phi} \left[ \frac{ \kappa_{x,n}^2+ \kappa_{z,n}^2  }{k_n^2 }\right] 
	-  \sum\limits_m   \tilde{\mu}_m \varPsi_{m+n}  =0,
	\label{eqa:recurs_gen}
\end{equation}
\end{subequations}
where $\tilde{\mu}_m=1/a_m$ with $a_m$s given by Eq.~\eqref{eqa:a_n}. Truncating to $2N+1$ terms yields  
\begin{subequations}
	\begin{equation}\label{eqa:Equation}
	[\mathbf{A}] \cdot [\vec{\Psi}] = 0,
	\end{equation}
where the matrix $[\textbf{A}]$ is a square matrix of size $(2N+1) \times (2N+1)$, that is,
\begin{equation}\label{eqa:matr}
		[\textbf{A}]=	\begin{bmatrix}
			v_{-N}  &  \tilde{\mu}_1 & \tilde{\mu}_2& \cdots  & \tilde{\mu}_{M-2}&\tilde{\mu}_{M-1}&\tilde{\mu}_M\\
			\tilde{\mu}_{-1}  &  v_{-N+1} & \tilde{\mu}_{1} &\cdots & \tilde{\mu}_{M-3}&\tilde{\mu}_{M-2}&\tilde{\mu}_{M-1}\\
			\tilde{\mu}_{-2}&\tilde{\mu}_{-1} &  v_{-N+2} & \cdots & \tilde{\mu}_{M-4}&\tilde{\mu}_{M-3}&\tilde{\mu}_{M-2}\\
			\vdots        &    \vdots    &    \vdots  &    \ddots  & \vdots  &        \vdots          &      \vdots\\
			\tilde{\mu}_{-M+2} 	&\tilde{\mu}_{-M+3} & \tilde{\mu}_{-M+4} &\cdots &	 v_{N-2} & \tilde{\mu}_{1}&\tilde{\mu}_{2}\\
			\tilde{\mu}_{-M+1} 	&\tilde{\mu}_{-M+2} & \tilde{\mu}_{-M+3} &\cdots &	\tilde{\mu}_{-1}  &  v_{N-1} & \tilde{\mu}_{1}\\
			\tilde{\mu}_{-M}  &  \tilde{\mu}_{-M+1} & \tilde{\mu}_{-M+2} &\cdots   &	\tilde{\mu}_{-2}&\tilde{\mu}_{-1} & v_{N}\\
		\end{bmatrix},
\end{equation}
where
\begin{equation}
v_n=\tilde{\mu}_0- \dfrac{\kappa_{x,n}^2+ \kappa_{z,n} ^2}{k_n^2}.
	\label{eqa:v}
\end{equation}
\indent The vector of unknowns $[\overrightarrow{\varPsi}]$ in Eq.~\eqref{eqa:Equation} is a $(2N+1)\times 1$ vector containing the $\varPsi_n$ values. For non-trivial solutions (i.e., $[\overrightarrow{\varPsi}] \neq 0$), the matrix $[\textbf{A}]$ must be singular, which implies that its determinant is zero, i.e.,
\begin{equation}
	\det[\mathbf{A}(\omega_n, \kappa_{x,n})] =0.
	\label{eqa:det_gen}
\end{equation}
\end{subequations}
\indent The dispersion relation condition in Eq.~\eqref{eqa:det_gen} ensures that the system supports wave propagation, and provides the dispersion relation, which is typically expressed as $\omega_n(\kappa_{x,n})$ or $\kappa_{x,n}(\omega_n)$. Equation~\eqref{eqa:det_gen}, along with the truncated $a_n$ coefficients (noting that, theoretically, an infinite number of elements exists), demonstrates the impact of nonlinearity, where energy from the primary harmonic modes ($A_{nn}$) is distributed across an infinite series of adjacent harmonic modes ($A_{nm}$, where $n \neq m$, $-\infty < n, m < +\infty$). This contrasts with linear space-time-modulated media~\cite{Taravati_PRAp_2018,taravati_PRApp_2019,Taravati_Kishk_TAP_2019}, where energy from the main harmonic modes ($A_{nn}$) is distributed only to the immediately adjacent harmonic modes ($A_{n,n-1}$ and $A_{n,n+1}$). Nonlinear systems, however, possess the ability to redistribute energy across a broader spectrum of harmonic frequencies, allowing efficient transfer of energy from the incident wave to higher-frequency harmonic modes. This is because the nonlinear interactions can sustain the high-frequency modulation necessary for such transfers. 

Consider an electromagnetic wave incident from free space (\(x < 0\)) at frequency \(\omega_0\) and incidence angle \(\theta_i\). The incident wavevector components are
\begin{subequations}
\begin{equation}
k_{x,\rm i} = k_0 \cos\theta_i, \qquad k_{z,\rm i} = k_0 \sin\theta_i,
\end{equation}
where \(k_0 = \omega_0 / c\). The magnetic and electric fields of the incident wave read
\begin{equation}
	\mathbf{H}_\text{I} (x,z,t)= \mathbf{\hat{y}} H_0 e^{j \omega_0 t} e^{-j\left[k_0\cos(\theta_\text{i}) x +k_0 \sin(\theta_\text{i}) z \right]},
\end{equation}
\begin{equation}
	\begin{split}
	&	\mathbf{E}_\text{I} (x,z,t)= -\eta_1 \left[\mathbf{\hat{k}}_\text{I} \times \mathbf{H}_\text{I} (x,z,t)\right] = \eta_1 \large[\mathbf{\hat{x}} \sin(\theta_\text{i}) \\&-\mathbf{\hat{z}} \cos(\theta_\text{i}) \large]     H_0 e^{j \omega_0 t} e^{-j\left[k_0\cos(\theta_\text{i}) x +k_0 \sin(\theta_\text{i}) z  \right]},
	\end{split}
\end{equation}
\end{subequations}
where $\eta_1=\sqrt{\mu_0 \mu_\text{r}/(\epsilon_0\epsilon_r)}$. For the reflected \(n\)-th harmonic, which has frequency \(\omega_0 + n\omega_s\) and free-space wavenumber \(k_{0n} = (\omega_0 + n\omega_s)/c\), the phase matching condition at the interface \(x = 0\) requires continuity of the tangential wavevector component for every Floquet harmonic \(n\), that is,
\begin{subequations}\label{eqa:A-ER_ET_forw}
\begin{equation}
k_{z,n}^{\rm R} = \kappa_{z,n} = k_0 \sin\theta_i + n \kappa_s,
\end{equation}
and the $x$-component and the unit wavevectoor of the reflected fields read
\begin{equation}
	k_{x,n}^{\rm R} = - k_{0n} \cos\theta_n^{\rm R}= - k_{0n} \sqrt{1 - \sin^2\theta_n^{\rm R}},
\end{equation}
\begin{equation}
	\mathbf{\hat{k}}^\text{R}_n=-\mathbf{\hat{x}} \cos(\theta^\text{R}_n) +\mathbf{\hat{z}} \sin(\theta^\text{R}_n).
\end{equation}
\indent The angle of the reflected harmonics is given by
\begin{equation}
\sin\theta_n^{\rm R} = \frac{k_{z,n}^{\rm R}}{k_{0n}} = \frac{k_0 \sin\theta_i + n \kappa_s}{k_{0n}},
\end{equation}
and the reflected fields outside of the slab read
	\begin{equation}
		\mathbf{H}_\text{R} (x,z,t)= \mathbf{\hat{y}} \sum\limits_{n =  - \infty }^\infty  R_{n} e^{-j \left[ -k_{0n} \cos(\theta_n^\text{R}) x +k_{0n} \sin(\theta_n^\text{R}) z  -\omega_n t\right] } ,
		\label{eqa:H_refl}
	\end{equation}
	\begin{equation}
		\begin{split}
			&\mathbf{E}_\text{R} (x,z,t)= -\eta_1 [\mathbf{\hat{k}}^\text{R}_n  \times \mathbf{H}_\text{R} (x,z,t)]= \eta_1 \large[\mathbf{\hat{x}} \sin(\theta^\text{R}_n) \\ &+\mathbf{\hat{z}} \cos(\theta^\text{R}_n) \large]    R_{n}   e^{-j \left[ k_{0} \cos(\theta_{\text{i}}) x -k_{0n} \sin(\theta^\text{R}_n) z  -\omega_n t\right] }.
			\label{eqa:E_refl}
		\end{split}
	\end{equation}
\end{subequations}
\indent We then enforce the continuity of the tangential components of the electromagnetic fields at $x=0$ to find the unknown field amplitudes $\varPsi_{n0}$ and $R_{n}$. The electric and magnetic fields continuity conditions at $x=0$, ${H_{1y}}(0,z,t) = {H_{2y}}(0,z,t)$ and ${E_{1z}}(0,z,t) = {E_{2z}}(0,z,t)$ yield
\begin{subequations}
\begin{equation}\label{eqa:EBC_forw_12}
	 R_{n} =    \varPsi_{n}-\delta_{n0} H_0,
\end{equation}
and
\begin{equation}\label{eqa:HBC_forw_12}
	\begin{split}
		& \cos(\theta_\text{i})	\delta_{n0} H_0  -\cos(\theta^\text{R}_n) R_{n}   =  \dfrac{\eta_2}{\eta_1}  \cos(\theta_n) \varPsi_{n},
	\end{split}
\end{equation}
which yields
\begin{equation}\label{eqa:HBC_forw_23}
	\begin{split}
	\varPsi_{0}= H_0\dfrac{ \cos(\theta_\text{i})+ \cos(\theta^\text{R}_0) }{ \cos(\theta_0) \eta_2/\eta_1+\cos(\theta^\text{R}_0) }.
	\end{split}
\end{equation}
\end{subequations}
\indent The procedure for determining the reflected fields is as follows. First, Eq.~\eqref{eqa:det_gen} is solved to obtain \(\kappa_{x,n}\). Second, Eq.~\eqref{eqa:HBC_forw_23} is solved for \(\Psi_0\), after which Eq.~\eqref{eqa:Equation} is solved for all \(\Psi_n\). The reflection coefficients \(R_n\) are then computed from Eq.~\eqref{eqa:EBC_forw_12}. Finally, the reflected magnetic and electric fields are obtained from Eqs.~\eqref{eqa:H_refl} and~\eqref{eqa:E_refl}, respectively.

\section{Results}~\label{sec:res}
\begin{figure}
	\begin{center}
			\includegraphics[width=1\columnwidth]{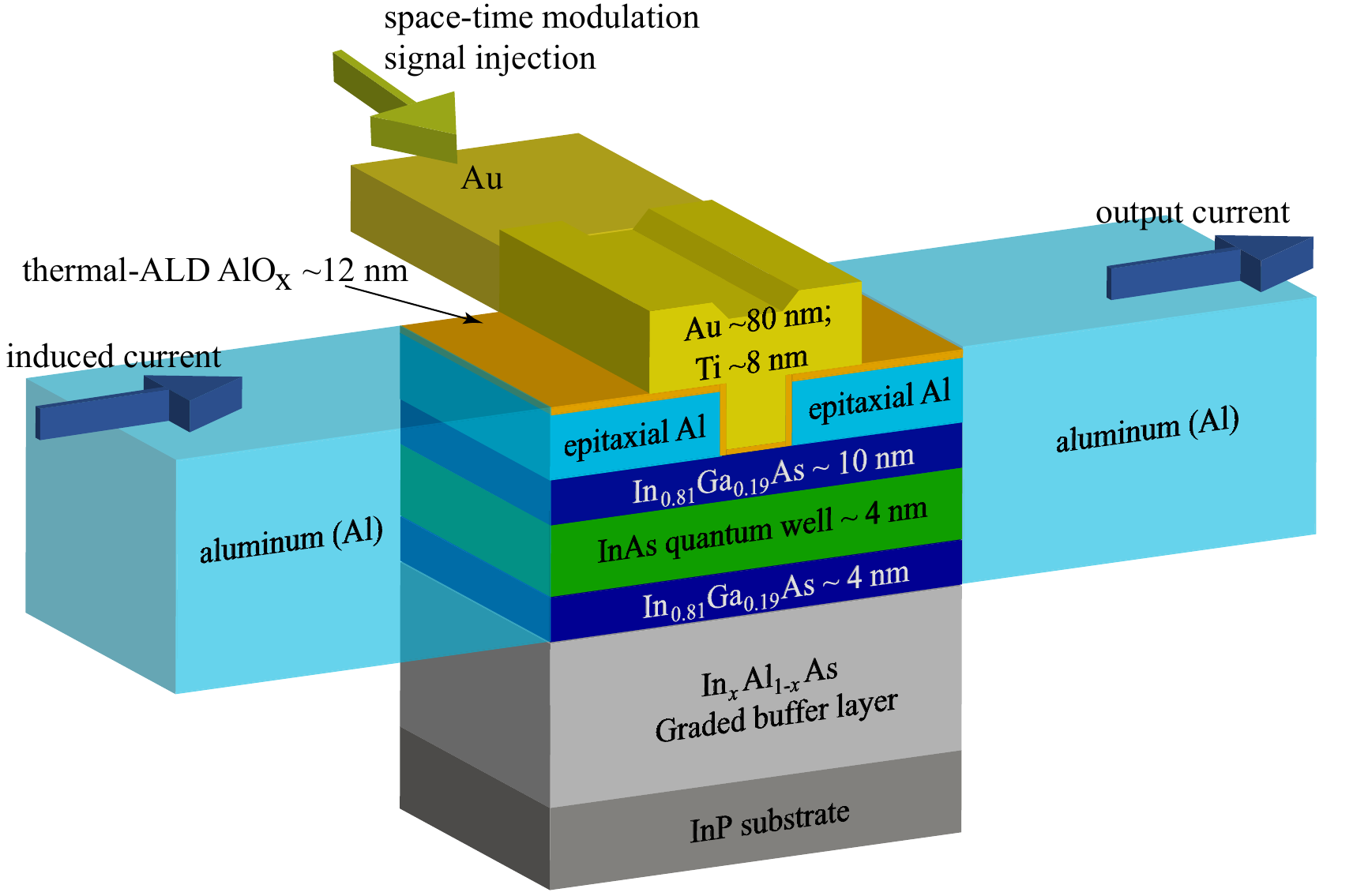}
		\caption{Schematic of the space-time-modulated superconductor-semiconductor Josephson field-effect transistor (JoFET).}
		\label{Fig:unitcell}
	\end{center}
\end{figure}
To demonstrate the functionality of the proposed nonreciprocal superconducting metasurface, we design a metasurface with thickness $d=0.6 \lambda_0$ where $\lambda_0=2\pi c/\omega_0$ is the free-space wavelength, $c$ is the speed of light in vacuum, and $\omega_0=\omega_\text{s}$ is the angular frequency of the incident wave. The space-time-varying permeability given by $\mu_\text{s}(z,t)= \sec\left(0.6+ 0.75 \sin[2\pi\times4\times 10^{9}(1.67 z/c- t)]\right)$. The choice of \(\mu_\text{s}(z,t) = \sec(0.6 + 0.75 \sin[2\pi \times 4 \times 10^{9}(1.67 z/c - t)])\) is guided by both the Josephson physics and the requirements for nonreciprocal photon-blockade analogue absorption. The secant form arises directly from the Josephson inductance of the JoFET array. The DC bias \(\widetilde{\Phi}_\text{dc} = 0.6\) ensures operation in a stable nonlinear regime, while the RF amplitude \(\widetilde{\Phi}_\text{rf} = 0.75\) provides strong harmonic coupling. The modulation frequency \(\omega_\text{s}/(2\pi) = 4\) GHz matches the incident frequency \(\omega_0\) for temporal coherence and is compatible with superconducting quantum circuits. The spatial wavenumber \(\kappa_s = 2\pi \times 4 \times 10^9 \times 1.67/c\) sets a subluminal modulation velocity, leading to the convergence of forward harmonics in the band structure and enabling directional absorption. Figure~\ref{Fig:unitcell} illustrates the experimental space-time-modulated JoFET-based unit-cell design. Space-time modulation is achieved through a periodic signal traveling along the yellow lines and spatiotemporally modulate the gate voltage of JoFETs. The proposed unit cell utilizes an Al-proximitized InAs quantum well, grown on a semi-insulating Fe counter-doped (100) InP wafer, providing a voltage-controlled Josephson junction~\cite{mayer2020gate,o2021epitaxial,phan2023gate}.

\begin{figure}
	\begin{center}
		\subfigure[]{\label{Fig:dispersion}
			\includegraphics[width=1\columnwidth]{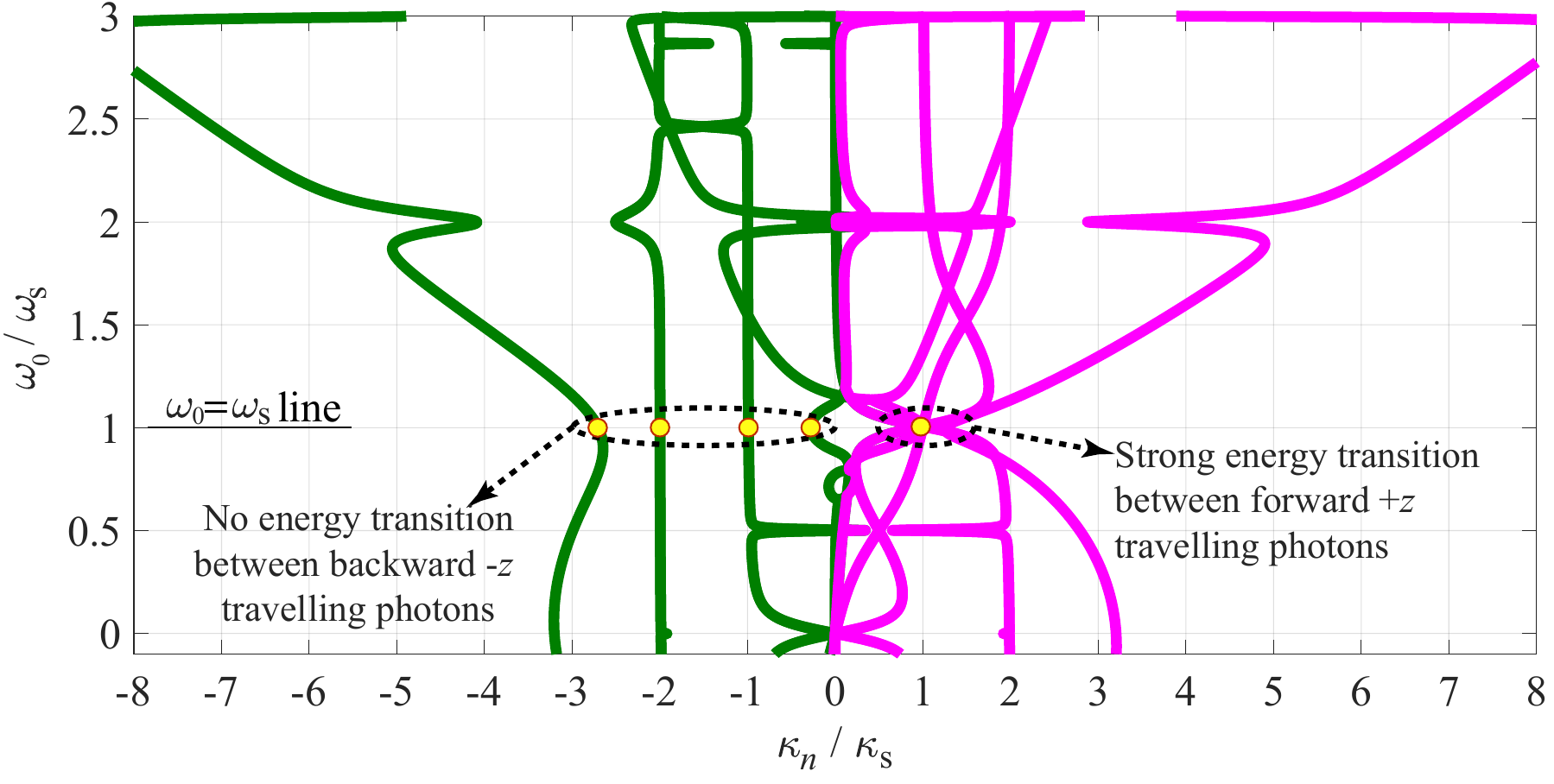}}
		\subfigure[]{\label{Fig:isof}
			\includegraphics[width=1\columnwidth]{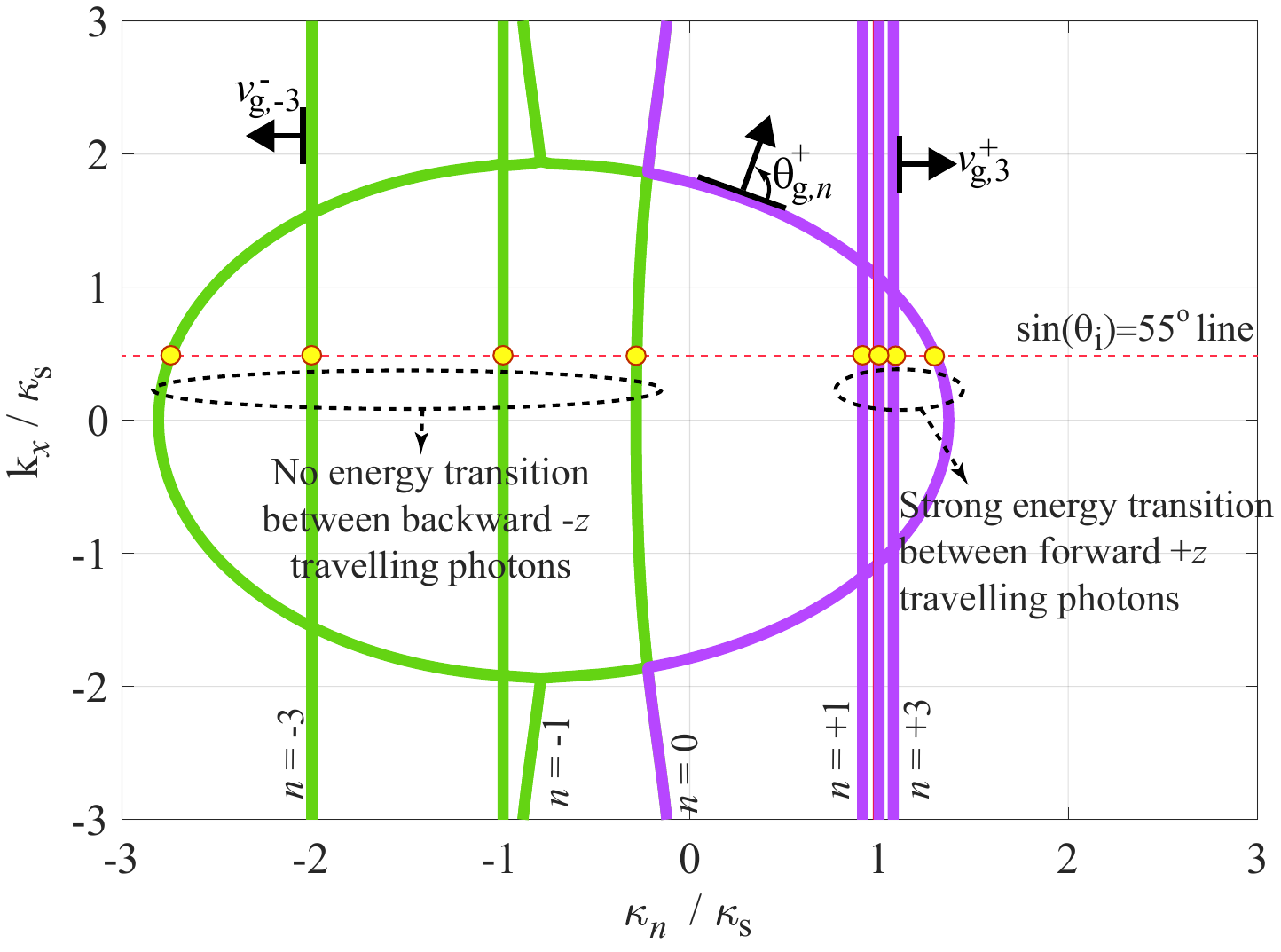}}
		\caption{Band structure of the superconductor-semiconductor spatiotemporal quantum metasurface computed using Eq.~\eqref{eqa:det_gen}. (a) The $\omega - \kappa$ diagram illustrates strong nonreciprocity at $\omega_0/\omega_\text{s} = 1$. Here, all higher-order forward ($+z$ direction) traveling space-time harmonics converge at $\kappa_n/\kappa_\text{s} = 1$, resulting in the absorption of energy from the fundamental harmonic and transitioning the system from $\omega_0$ to higher frequency states, denoted as $\omega_m$. In contrast, the higher-order backward ($-z$ direction) traveling space-time harmonics are spatially separated, preventing any energy transition from $\omega_0$ to higher states. (b) The $k_x - \kappa_n$ isofrequency diagram at $\omega_0/\omega_\text{s} = 1$ further demonstrates the nonreciprocity of the metasurface, where all forward higher-order harmonics are clustered at $\kappa_n/\kappa_\text{s} = 1$, effectively absorbing the energy of the fundamental harmonic. The group velocity vector $v_{\text{g},n}$ indicates that these harmonics propagate along the $+z$ direction, parallel to the metasurface boundary, rather than being transmitted outside the metasurface.}
		\label{Fig:res}
	\end{center}
\end{figure}

To understand the nonreciprocal behavior of the superconductor-semiconductor spatiotemporal quantum metasurface, we first analyze its band structure using the dispersion relation. The analysis reveals crucial insights into the energy transitions and nonreciprocity induced by the space-time modulation. Figure~\ref{Fig:dispersion} presents the $\omega - \kappa$ diagram, showcasing the band structure of the metasurface at the critical point where $\omega_0/\omega_\text{s} = 1$. This diagram illustrates a strong nonreciprocity in the system. Specifically, the higher-order harmonics that travel in the forward $+z$ direction converge at the normalized wave vector $\kappa_n/\kappa_\text{s} = 1$. This convergence indicates a resonant interaction between the incident wave and the space-time-modulated medium, leading to the absorption of energy from the fundamental harmonic. The absorbed energy causes a transition from the fundamental frequency $\omega_0$ to higher energy states denoted as $\omega_m$.

In contrast, the higher-order harmonics that travel in the backward $-z$ direction are spatially separated in the $\omega - \kappa$ diagram. This separation prevents any significant interaction between the incident wave and the medium, thereby inhibiting the energy transition from $\omega_0$ to higher states. As a result, the backward-propagating waves do not experience the same degree of energy absorption as the forward-propagating waves, which directly contributes to the observed nonreciprocal behavior of the metasurface. Figure~\ref{Fig:isof} shows the $k_x - \kappa_n$ isofrequency diagram at $\omega_0/\omega_\text{s} = 1$. This diagram further confirms the nonreciprocal nature of the metasurface, where all forward-propagating higher-order harmonics are clustered at $\kappa_n/\kappa_\text{s} = 1$. The clustering of these harmonics facilitates the absorption of the fundamental harmonic's energy, while the group velocity vector $v_{\text{g},n}$ indicates that the energy is predominantly confined and directed along the $+z$ axis, parallel to the metasurface boundary. This confinement prevents the transmission of energy to the exterior of the metasurface, leading to strong nonreciprocal absorption.

The phenomena observed in the band structure is related to the concept of spatiotemporal photon blockade, where the presence of a single photon can prevent the passage of subsequent photons due to nonlinear interactions, effectively "blocking" further photon transmission. In our system, this effect is realized through the spatiotemporal modulation of the metasurface. The space-time modulation, with a frequency $\omega_\text{s}$ matched to the incident photon frequency $\omega_0$, induces a strong nonlinear interaction for forward-propagating waves. As a result, when photons from the left are incident on the metasurface, they are resonantly coupled to the modulation, leading to energy absorption and transition to higher frequency states (spatial harmonics) $\omega_m$. This absorption effectively prevents the transmission through the metasurface. The nonreciprocity arises because the same modulation does not facilitate similar energy transitions for backward-propagating waves, allowing them to pass through the metasurface without significant energy loss. Thus, the metasurface behaves as a nonreciprocal device, where the space-time modulation selectively blocks traveling in one direction by shifting their energy to higher states, while allowing wave traveling in the opposite direction to pass freely. This behavior is crucial for potential applications in quantum information processing and nonreciprocal quantum devices, where controlling the directionality and energy states of photons is essential.

\subsection{Transmission and fields distributions}
\begin{figure*}
	\begin{center}
		\subfigure[]{\label{Fig:Ey_time}
			\includegraphics[width=0.9\columnwidth]{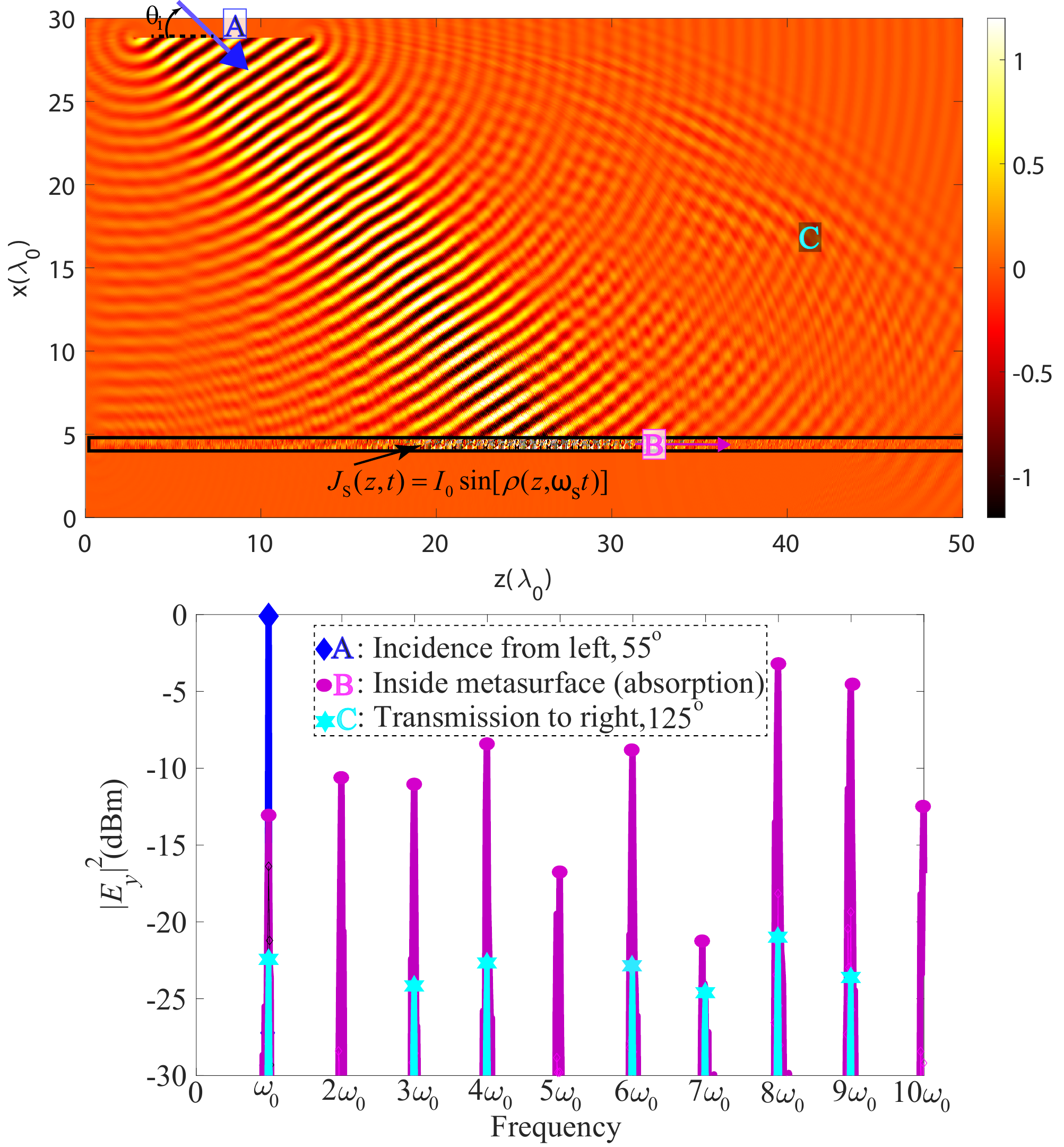}}
		\subfigure[]{\label{Fig:Ey_spec}
			\includegraphics[width=0.9\columnwidth]{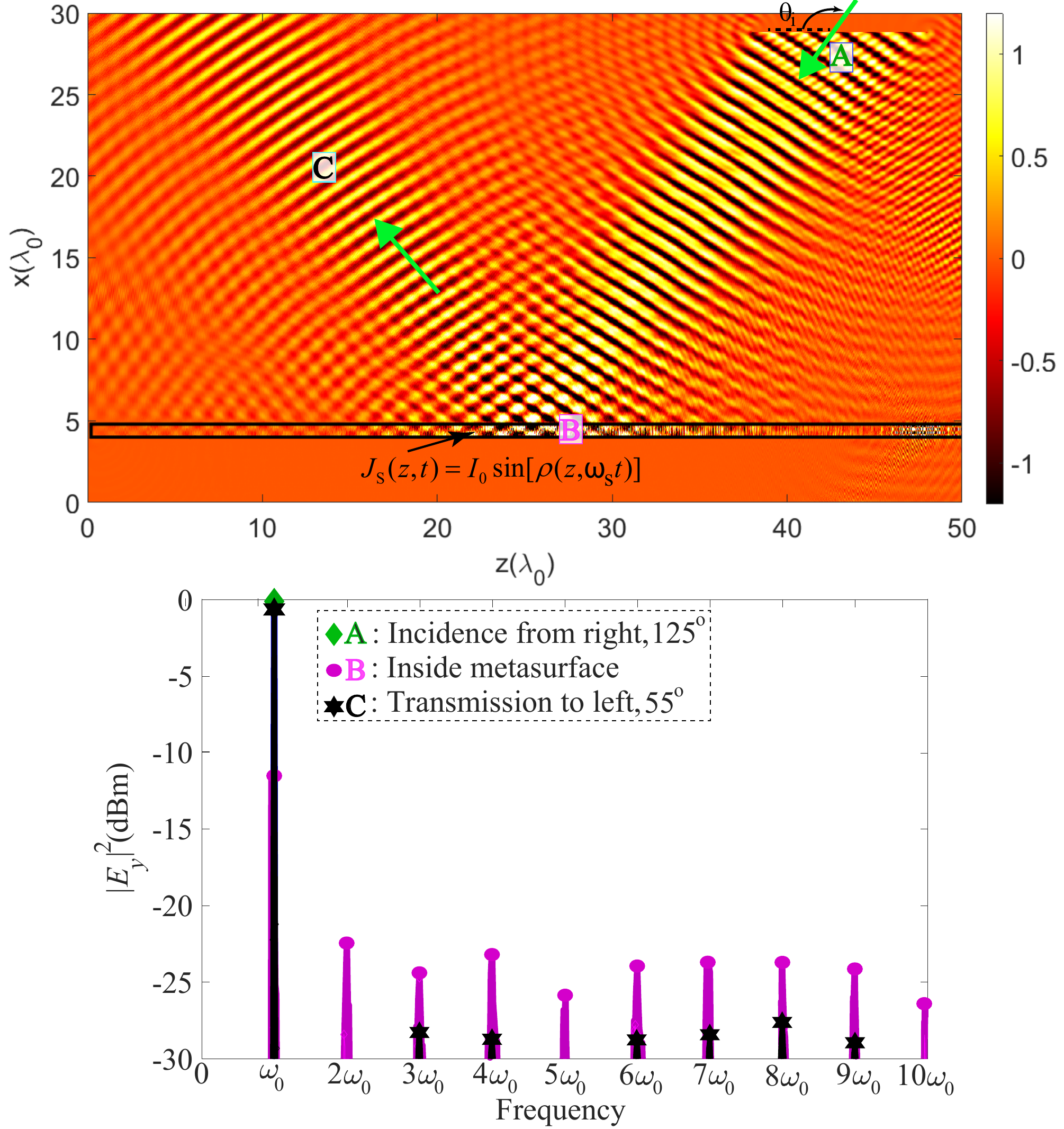}}\\
		\subfigure[]{\label{Fig:Hz_time}
			\includegraphics[width=0.9\columnwidth]{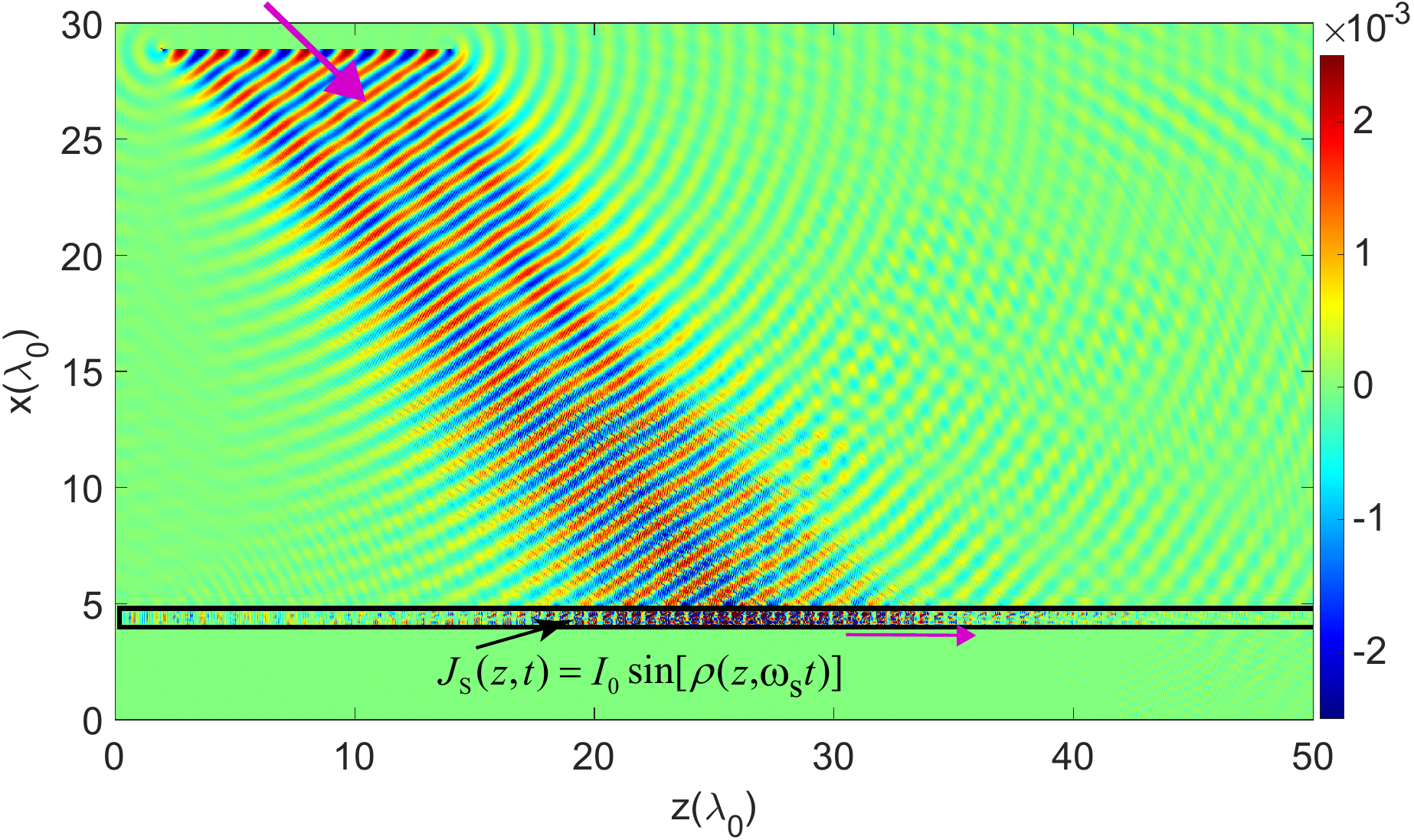}}
		\subfigure[]{\label{Fig:Hz_spec}
			\includegraphics[width=0.9\columnwidth]{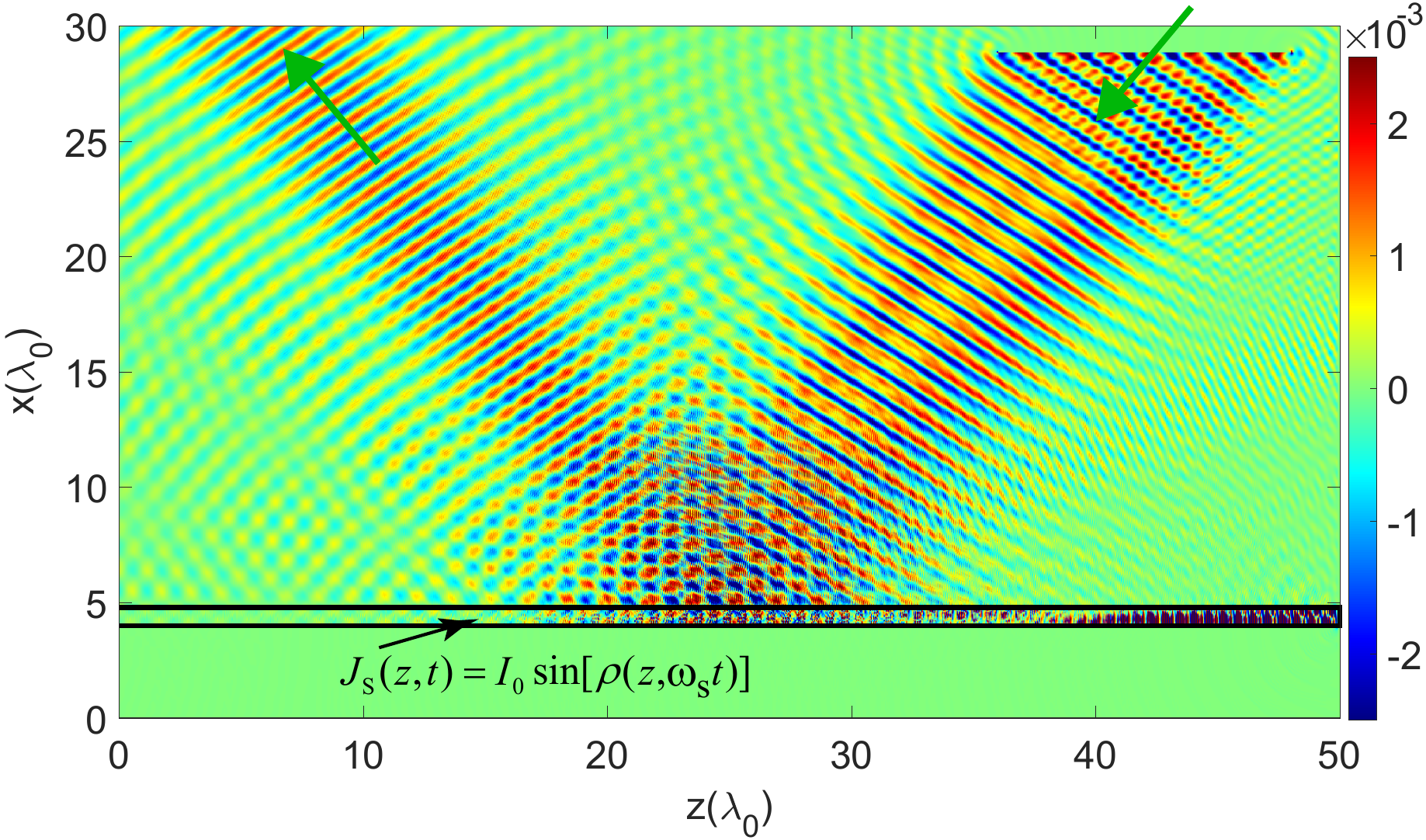}}
		\caption{Nonreciprocal absorption in spatiotemporal superconductor-semiconductor quantum metasurfaces. (a) and (b) Field distribution (top) and frequency spectrum (bottom) for $E_y$, demonstrating (a) strong absorption of the incident beam from the left at $55^\circ$, and (b) full transmission of the incident beam from the right at $125^\circ$ to the left at $55^\circ$. (c) and (d) Field distribution for $H_z$, demonstrating (c) strong absorption of the incident beam from the left at $55^\circ$, and (d) full transmission of the incident beam from the right at $125^\circ$ to the left at $55^\circ$.}
		\label{Fig:result}
	\end{center}
\end{figure*}
Next, we analyze the nonreciprocal absorption characteristics of the spatiotemporal superconductor-semiconductor quantum metasurface. The metasurface is designed to exhibit strong nonreciprocal behavior due to the space-time modulation, leading to direction-dependent absorption and transmission properties. Figures~\ref{Fig:Ey_time} and~\ref{Fig:Ey_spec} illustrate the electric field distribution~($E_y$) and corresponding frequency spectrum for two different incident angles. Specifically, Figure~\ref{Fig:Ey_time} demonstrates the field distribution and frequency response when an electromagnetic wave is incident from the left at an angle of $55^\circ$. The results clearly show strong absorption of the incident wave by the metasurface, with minimal transmission. The corresponding frequency spectrum confirms the absorption by showing the lack of transmitted energy at the incident frequency. Conversely, Figure~\ref{Fig:Ey_spec} presents the scenario where the incident beam approaches from the right at an angle of $125^\circ$. In this case, the metasurface allows full transmission of the wave to the left side, effectively reflecting the wave back towards the source at an angle of $55^\circ$. The frequency spectrum corroborates this by displaying a significant transmitted signal at the original frequency, confirming the nonreciprocal behavior of the metasurface. Figures~\ref{Fig:Hz_time} and~\ref{Fig:Hz_spec} provide the magnetic field distribution ($H_z$) under the same conditions. Figure~\ref{Fig:Hz_time} shows the absorption of the left-incident wave at $55^\circ$, while Figure~\ref{Fig:Hz_spec} demonstrates the full transmission of the right-incident wave at $125^\circ$. The consistent behavior observed in both the electric and magnetic field distributions confirms the robustness of the nonreciprocal absorption mechanism induced by the spatiotemporal modulation in the superconducting quantum metasurface. These results demonstrate the capability of the designed metasurface to enforce nonreciprocity, effectively controlling the directionality of absorption and transmission. The strong interaction between the incident photons and the spatiotemporally modulated nonlinear metasurface unit cells underpins the observed nonreciprocal behavior, which could be leveraged in various millikelvin-temperature quantum technology applications requiring directional control of wave propagation.

\section{Conclusions}
we have introduced and theoretically demonstrated a spatiotemporally modulated superconducting metasurface that exhibits photon-blockade-analogue nonreciprocal absorption. By matching the modulation frequency to the incident wave frequency (\(\omega_\text{s} = \omega_0\)), we have shown that forward-traveling waves undergo resonant coupling to higher-order Floquet harmonics, leading to strong absorption within the metasurface, while backward-traveling waves transmit freely without significant interaction. This directional behavior arises from classical harmonic conversion in a space-time periodic medium, constituting a classical analogue of quantum photon blockade. We have derived a comprehensive theoretical framework encompassing the system Hamiltonian, Floquet band structure, and full-wave simulations. We have derived a first-principles theoretical framework starting from the microscopic Hamiltonian of a single JoFET cell, $H_j=4E_Cn_j^2-E_J(V_g,t)\cos\phi_j$. The classical Josephson relations, the gate-driven Josephson energy, and the array's effective space-time-periodic permeability all follow directly from this Hamiltonian via its Heisenberg equations of motion, rather than being imposed independently, before quantization and connection to the classical wave equation. The band structure analysis revealed that the convergence of forward-propagating harmonics is the key mechanism enabling nonreciprocal absorption, while backward-propagating harmonics remain spatially separated and uncoupled. Full-wave simulations confirmed strong directional absorption with minimal transmission for left-incident waves and full transmission for right-incident waves. The proposed metasurface, based on an Al-InAs JoFET array, is compatible with millikelvin-temperature superconducting quantum technologies and offers a compact, magnet-free solution for directional absorption. Unlike conventional nonreciprocal devices such as circulators and isolators, which require bulky ferrite materials or external magnetic fields, our approach achieves nonreciprocity through spatiotemporal modulation alone, making it suitable for on-chip integration in superconducting quantum circuits. These findings establish a pathway toward a new class of nonreciprocal superconducting devices for quantum information processing, microwave photonics, and photon management in quantum systems. Future work may explore the extension of this concept to other frequency regimes, the incorporation of active gain elements for nonreciprocal amplification, and the experimental realization of the proposed metasurface in cryogenic environments.

%%%%%%%%%%%%%%%% REFERENCES %%%%%%%%%%%%%%%

%\clearpage % Clear all remaining figures and tables then start a new page

% The list of references goes after the main text and before the acknowledgements
% When preparing an initial submission, we recommend you use BibTeX, like this:
%
\bibliography{Taravati_Reference}
%\printbibliography

\end{document}